\documentstyle[12pt]{article}
\def\be{\begin{eqnarray}}
\def\ee{\end{eqnarray}}
\def\ba{\begin{array}}
\def\ea{\end{array}}

\begin{document}

\begin{center}
{\bf\Large 
{Bosonic string \,--\, Kaluza Klein theory\\
\vskip 3mm
exact solutions using 5D--6D dualities}}
\end{center}

\vskip 1cm

\begin{center}
{\bf \large {Alfredo Herrera-Aguilar}}
\end{center}

\begin{center}
Instituto de Fisica y Matematica,\\
Universidad Michoacana AP 2-82, CP 58040,\\
Morelia, Michoacan, Mexico\\
e-mail: herrera@zeus.ccu.umich.mx
\end{center}

\vskip 3mm
\begin{center}
and
\end{center}
\vskip 5mm

\begin{center}
{\bf \large {Oleg V. Kechkin}}
\end{center}

\begin{center}
Institute of Nuclear Physics,\\
M.V. Lomonosov Moscow State University, \\
Vorob'jovy Gory, 119899 Moscow, Russia, \\
e-mail: kechkin@depni.npi.msu.su
\end{center}

%%%%%%%%%%%%%%%%%%%%%%%%%%%%%%%%%%%%%%%%%%%%%%%%%%%%%%%%%%%%%%%%%%%%%%%%%%%%%
\vskip 1cm
\begin{abstract}
We present the explicit formulae which allow to transform the general
solution of the $6D$ Kaluza--Klein theory on a $3$--torus into the 
special solution of the $6D$ bosonic string theory on a $3$--torus
as well as into the general solution of the $5D$ bosonic string theory on a
$2$--torus. We construct a new family of the extremal solutions of the
$3D$ chiral equation for the $SL(4,R)/SO(4)$ coset matrix and interpret it 
in terms of the component fields of these three duality related theories.
\end{abstract}
%%%%%%%%%%%%%%%%%%%%%%%%%%%%%%%%%%%%%%%%%%%%%%%%%%%%%%%%%%%%%%%%%%%%%%%%%%%%%%

\newpage

%%%%%%%%%%%%%%%%%%%%%%%%%%%%%%%%%%%%%%%%%%%%%%%%%%%%%%%%%%%%%%%%%%%%%%%%%%
\section{Introduction}

In the classical field theory it is possible a situation when the different
theories possess the same dynamics. This fact means that one can express 
the fields of one theory in terms of the fields of another theory by the
help of motion equations or without this help. In the first case one deals
with the on--shell equivalent theories, whereas in the second case this
equivalence has the more general off--shell type.

The string theories \cite{gsw} provide many examples of the field theories 
with 
equivalent dynamics. In the framework of (super)string theory the
dynamical equivalence is considered as some type of symmetry, or duality,
which acts between the different special string theories as between the
different special limits of the single general string theory \cite{kir}. 
However, from the classical point of view, the presence of such a
duality means the possibility to obtain new constructive information about 
one
theory from the known information about its dual theory. For 
example,
one can rewrite the exact solution of one theory in the form of exact 
solution of
its dual ``partner'' using the corresponding algebraical and differential 
relations which express this duality.
The differential relations become necessary in the case of the on--shell
equivalence, whereas for the off--shell equivalent theories one needs only
in the algebraical calculations.

In this work we present the explicit formulae which establish the off--shell
equivalence between the full Kaluza--Klein theory \cite {kk} coupled to 
dilaton and some
special subset of the bosonic string theory toroidally compactified to $3$ 
dimensions.
The special subset of the bosonic string thery is
defined with the zero value of extra Kalb--Ramond and mixed metric field 
components;
the number of the compactified dimensions is left arbitrary. In the special
case of the $6D$ theories we show that 
the pure Kaluza--Klein theory and the corresponding special subset of the 
bosonic string theory are on--shell equivalent, or dual, to
the full $5D$ bosonic string theory and give the explicit formulae which 
express this duality (see also \cite{y} and references therein). 
We show that these three theories are equivalent on--shell to
the chiral theory with the coset $SL(4,\,R)/SO(4)$ chiral matrix 
minimally coupled to the $3D$ General Relativity. We construct a new class
of the extremal solutions of this effective $3D$ theory and, using the dual 
relations,
calculate the corresponding solutions of the $6D$ and $5D$ theories under
consideration. The constructed solutions have the Majumdar--Papapetrou type
\cite{extr} and provide a good illustration of the established dualities 
because 
they possess the nontrivial values of all the essential duality related 
field components.

%%%%%%%%%%%%%%%%%%%%%%%%%%%%%%%%%%%%%%%%%%%%%%%%%%%%%%%%%%%%%%%%%%%%%%%%%%

\section{$\sigma$--model for bosonic string theory}

At low energies the bosonic string theory can be described by the
field theory of its massless modes. These modes live in $D$--dimensional 
space--time and include the dilaton field  $\Phi$, the Kalb--Ramond field 
$B_{MN}=-B_{NM}$ and the metric field $G_{MN}=G_{NM}$. The corresponding
action reads \cite{gsw}:
\be\label{1}
S_{D}=\int d^{D}X\sqrt{|{\rm det}\,G_{MN}|}e^{-\Phi} 
\left ( R_{D}+\Phi_{,M}\Phi^{,M}-\frac{1}{12}H_{MNK}H^{MNK}
\right ),
\ee
where $H_{MNK}=\partial_MB_{NK}+\partial_NB_{KM}+\partial_KB_{MN}$.

Below we consider the toroidal compactification of the first $d$ space--time
dimensions to the $3$--dimensional space, i.e., we put $D=d+3$ and consider
the fields independent on the coordinate $X^M$ with $M=m=1,...,d$ 
(below we denote $y^m\equiv X^m$) and possessing the
functional dependence on the coordinates $x^{\mu}\equiv X^{d+\mu}$ with
$\mu=1,2,3$. In this case the field components can be separated in respect to
the transformations of the three--dimensional coordinates $x^{\mu}$
in the following way (\cite{sen3}):

\vskip 3mm

\noindent
$1)$ \, two scalar matrices $G$ and $B$ of the dimensions $d\times d$, 
with the components $G_{mk}$ and $B_{mk}$ correspondingly, and the 
scalar function 
\be \label{1'}
\phi=\Phi-{\rm ln }\sqrt{|{\rm det}\,G|}; 
\ee

\vskip 3mm

\noindent
$2)$ \, two vector columns $\vec A_{1}$ and $\vec A_{2}$ of the dimension
$d\times 1$; their components read:
\be \label{2}
(\vec A_1)_{m\mu}&=&(G^{-1})_{mk}G_{k,d+\mu},
\nonumber\\
(\vec A_2)_{m\mu}&=&B_{m,d+\mu}-B_{mn}(\vec A_1)_{n\mu};
\ee
and

\vskip 3mm

\noindent
$3)$ \, two tensor fields
\be \label{3}
h_{\mu\nu}&=&e^{-2\phi}\left [ G_{d+\mu,\,d+\nu}-G_{mk}
(\vec A_1)_{m\mu}(\vec A_1)_{k\nu}
\right ],
\nonumber\\
b_{\mu\nu}&=&B_{d+\mu,\,d+\nu}-B_{mk}(\vec A_1)_{m\mu}(\vec A_1)_{k\nu}-
\frac{1}{2}\left [(\vec A_1)_{m\mu}(\vec A_2)_{m\nu}-(\vec
A_1)_{m\nu}(\vec A_2)_{m\mu}\right ].
\ee
In three dimensions the field $b_{\mu\nu}$ is nondynamical; following 
\cite{sen3} one can put $b_{\mu\nu}=0$ without any contradiction with 
the motion 
equations. Moreover, in three dimensions it is possible to introduce 
pseudoscalar fields $u$ and $v$ accordingly the relations 
(\cite{sen3}--\cite{mep})
\be \label{4}
\nabla\times\vec A_1&=&e^{2\phi}G^{-1}
\left (\nabla u+B\nabla v\right ),
\nonumber                            \\
\nabla\times\vec A_2&=&e^{2\phi}G\nabla v-
B\nabla\times\vec A_1.
\ee
Finally, the bosonic string theory toroidally
compactified to three dimensions is equivalent on shell to the effective 
three--dimensional theory which describes the interacting scalar fields $G$, 
$B$ and $\phi$ and the pseudoscalar ones $u$ and $v$ coupled to 
the metric $h_{\mu\nu}$. The corresponding action reads:
\be \label{5}
S_3=\int d^3x \sqrt h\left \{ -R_3+L_3\right \},
\ee
where
\be \label{6} 
L_3=\frac{1}{4}{\rm Tr}\left ( J_G^2-J_B^2\right ) + (\nabla\phi)^2-
\frac{1}{2}e^{2\phi}\left [ (\nabla u+B\nabla v)^TG^{-1}(\nabla u+B\nabla v)
+\nabla v^TG\nabla v\right ], 
\ee
\be \label{7}
J_G=\nabla G\,G^{-1}, \qquad J_B=\nabla B\,G^{-1},
\ee
$h={\rm det}\,h_{\mu\nu}$ and the scalar curvature $R_3$ is constructed using 
the metric $h_{\mu\nu}$. This action describes the nonlinear $\sigma$--model 
coupled to the three--dimensional gravity. Below we use this $\sigma$--model
to describe three different special theories.

%%%%%%%%%%%%%%%%%%%%%%%%%%%%%%%%%%%%%%%%%%%%%%%%%%%%%%%%%%%%%%%%%%%%%%%%

\section{Duality in arbitrary dimension}

Working with the field theory which contains several algebraically 
independent fields,
one can study its special cases when some part of these fields is taken 
in the form of given functions. For example, one can try to put some fields
of the theory equal to their trivial values (zero, for example) and to study
the dynamics of the remaining fields. This approach leads to two possible 
essentially different situations: the resulting dynamics can be free or 
restricted by
the set of additional relations. These relations can arise as the on--shell 
consequence
of our fixation of the fields subset; in the special situation the 
corresponding restrictions satisfied identically on shell of the remaining
field equations. In this ``successful'' situation we call
the set of the remaining fields as the ``subsystem'' of the original theory.

The effective $3$--dimensional theory (\ref{5})--(\ref{7}) possesses the 
following two subsystems:

\vskip 3mm
\noindent
I)\,\, the subsystem with the trivial (zero) values of the fields
$B$ and $v$; its Lagrangian reads:
\be \label{8}
L_I=(\nabla\phi)^2+\frac{1}{4}{\rm Tr}J_G^2-\frac{1}{2}e^{2\phi}
\nabla u^TG^{-1}\nabla u.
\ee
For this subsystem $\vec A_2=0$, whereas
\be \label{9}
\nabla\times\vec A_1=e^{2\phi}G^{-1}\nabla u;
\ee
and

\vskip 2mm

\noindent
II)\,\, the subsystem with the trivial (zero) values of the fields
$B$ and $u$ with the Lagrangian
\be \label{10}
L_{II}=(\nabla\phi)^2+\frac{1}{4}{\rm Tr}J_G^2-\frac{1}{2}e^{2\phi}
\nabla v^TG\nabla v.
\ee
Here $\vec A_1=0$ and
\be \label{11}
\nabla\times\vec A_2=e^{2\phi}G\nabla v.
\ee

It is easy to see that the map
\be \label{12}
\phi\rightarrow\phi,\qquad G\rightarrow G^{-1},\qquad
u\rightarrow v
\ee
transforms the subsystem ``I'' into the subsystem ``II''. For the vector
variables this map gives 
\be \label{13}
\vec A_1\rightarrow \vec A_2.
\ee
To obtain the physical interpretation of these subsystems one must
use the definitions (\ref{1'})--(\ref{4}) of the $\sigma$--model 
(\ref{6}) variables. 
As the result one concludes that the subsystem ``I''
corresponds to the full Kaluza--Klein theory coupled to the dilaton field,
whereas the subsystem ``II'' describes the bosonic string theory without
extra components of the Kalb--Ramond field and without the mixed 
(``rotational'')
components of the metric. The map (\ref{12}) provides the equivalence of
the subsystems ``I'' and ``II''; this equivalence has the off--shell nature.
Actually, it is obviously for the effective $3$--dimensional theories 
described by Eqs. 
(\ref{8}) and (\ref{10}); however in fact we consider the physical theories
in $d+3$ dimensions. One must come back to the multidimensional fields to
see what kind of equivalence really takes place in this case. The map 
(\ref{12})--(\ref{13}) means the possibility to express the 
$(d+3)$--dimensional fields for the both subsystems in terms of the same 
symbols.
Let us take the subsytem ``I'' as the ``starting'' one; let us denote
the matrix $G$, the rotational vector $\vec A_1$ and the multidimensional 
dilaton $\Phi$ for this subsytem as $F$, $\vec\omega$ and $\Phi$ 
correspondingly. Then for the subsystem ``I'' one has:
\be \label{14}
ds^2_I&=&(dy+\vec\omega d\vec x)^TF(dy+\vec\omega d\vec x)+
\frac{e^{2\Phi}}{f}h_{\mu\nu}dx^{\mu}dx^{\nu},
\nonumber\\
\Phi_I&=&\Phi,
\ee
where
\be \label{15}
f=|{\rm det}\,F|.
\ee
Below we will need in the column $u$ for the subsystem ``I''; let us also put
\be \label{15'}
u_I=H.
\ee
Then, using the map (\ref{12})--(\ref{13}) and Eqs. (\ref{1'})--(\ref{4}),
for the subsystem ``II'' one immediately obtains the following expressions
for the nonzero field components:
\be \label{16}
ds^2_{II}&=&dy^TF^{-1}dy+
\frac{e^{2\Phi}}{f}h_{\mu\nu}dx^{\mu}dx^{\nu},
\nonumber\\
\Phi_{II}&=&\Phi-{\rm ln}\,f,\qquad B_{II\,\,\,m,\,d+\mu}=
(\vec\omega )_{m\mu}.
\ee
Thus, one can see that our subsystems are off--shell equivalent because 
Eqs. (\ref{14}) and (\ref{16}) are expressed in terms of the same variables
using only the algebraical (not differential) operations. In the following
section we will see that this fact is not trivial; in this section 
the subsytem ``III'',
related to the subsystems ``I'' and ``II'' via the differential as well as 
the algebraical relations, will be introduced. 

%%%%%%%%%%%%%%%%%%%%%%%%%%%%%%%%%%%%%%%%%%%%%%%%%%%%%%%%%%%%%%%%%%%%%%%%%%

\section{Duality between 6D and 5D theories}

Let us truncate the subsystems ``I'' and ``II'' by putting $\Phi=0$ in Eqs. 
(\ref{14}) and (\ref{16}). In this case the subsystem ``I'' describes the 
pure Kaluza--Klein theory, whereas the subsystem ``II'' possesses the 
multidimensional dilaton field uniquely related to the extra metric 
components. Now we would like to develop some useful matrix formalisms for 
these simplified subsystems. Namely, for the Kaluza--Klein theory the 
effective $3$--dimensional Lagrangian can be represented in the following 
(chiral) form \cite{slkk}:
\be \label{17}
L=\frac{1}{4}{\rm Tr}\, \vec J_{M}^2,
\ee
where $\vec J_{M}=\nabla M\,M^{-1}$. Here the matrices
$M$ and $M^{-1}$ are constructed from the $\sigma$--model fields 
$F$ and $H$ as
\be \label{18}
M=
\left(
\ba{cc}
F^{-1}&F^{-1} H\cr 
H^TF^{-1}&f+H^TF^{-1}H
\ea
\right),\quad
M^{-1}=
\left(
\ba{cc}
F+f^{-1}HH^T&-f^{-1} H\cr 
-f^{-1}H^T&f^{-1}
\ea
\right).
\ee
It is easy to see that the matrix $M$ satisfies the 
$SL(d+1,R)/SO(d+1)$ coset relations
\be \label{19}
{\rm det}\, M=1, \qquad M^T=M.
\ee
From the duality (\ref{12}) it follows that Eq. (\ref{17}) also gives 
the effective action for the simplified subsystem ``II''; in this case
$H\equiv v_{II}$.

Then, the effective $3$--dimensional Lagrangian (\ref{6}) of the bosonic 
string theory also can be represented in the chiral form 
\cite{sen3}--\cite{mep}:
\be \label{20}
L_{BS}=\frac{1}{8}{\rm Tr}\, \vec J_{{\cal M}_{BS}}^2,
\ee
where $\vec J_{{\cal M}_{BS}}=\nabla {\cal M}_{BS}\,{\cal M}_{BS}^{-1}$. 
Here the chiral matrix ${\cal M}_{BS}$ reads:
\be \label{21}
{\cal M}_{BS}=
\left(
\ba{cc}
{\cal G}^{-1}&{\cal G}^{-1} {\cal B}\cr 
-{\cal B}^T{\cal G}^{-1}&{\cal G}-{\cal B}^T{\cal G}^{-1}{\cal B}
\ea
\right),
\ee
and the block components ${\cal G}$,\, ${\cal G}^{-1}$ and {\cal B} 
are defined as
\be \label{22}
{\cal G}=
\left(
\ba{cc}
-e^{-2\phi}+v^TGv&v^TG\cr 
Gv&G
\ea
\right),\quad
{\cal G}^{-1}=
\left(
\ba{cc}
-e^{2\phi}&e^{2\phi}v^T\cr 
e^{2\phi}v&G^{-1}-e^{2\phi}vv^T
\ea
\right),%\nonumber
\ee
\be \label{22'}
{\cal B}=
\left(
\ba{cc}
0&-(u+Bv)^T\cr 
u+Bv&B
\ea
\right).
\ee
It is easy to prove that ${\cal M}_{BS}$ yields the following coset 
relations:
\be \label{23}
{\cal M}_{BS}^T{\cal L}{\cal M}_{BS}={\cal L}, \qquad {\cal M}_{BS}^T=M_{BS},
\ee
where
\be \label{24}
{\cal L}=
\left(
\ba{cc}
0&1\cr 
1&0
\ea
\right),
\ee
so ${\cal M}_{BS}\in O(d+1,d+1)/O(d+1)\times O(d+1)$.

Now let us consider the special case of $d=2$, i.e., the (complete) 
$5$--dimensional bosonic string theory. In \cite{ky} it was shown that this 
theory can be represented by the following chiral matrix of the 
Kaluza--Klein theory type:
\be \label{25}
M_{BS}=\frac{1}{({\rm det}\, {\cal G})^{\frac{1}{2}}}
\left(
\ba{cc}
{\cal G}&{\cal G}h\cr 
h^T{\cal G}&{\rm det}\, {\cal G}+h^T{\cal G}h
\ea
\right),
\ee
where
\be \label{25'}
h_m=\frac{1}{2}\epsilon_{mnk}{\cal B}_{nk}.
\ee
In terms of this matrix the Lagrangian (\ref{6}) obtains the form (\ref{17}) 
of the Kaluza--Klein theory living in $6$ dimensions
($M_{BS}$ is the $4\times 4$ matrix):
\be \label{26}
L_{BS}=\frac{1}{4}{\rm Tr}\, \vec J_{M_{BS}}^2.
\ee
Using this fact it is possible to identify the $5D$ bosonic string theory 
with the $6D$ Kaluza--Klein theory, i.e., with the simplified subsystem
``I'' (and ``II'') in the case of $d=3$. The identification relations  
follow from the comparison of Eqs. (\ref{18}) and (\ref{25}):
\be \label{27}
{\cal G}=fF^{-1}, \qquad h=H.
\ee
In the remaining part of this paper we call the (complete) $5D$ bosonic 
string theory as the ``subsytem III''. Thus, the subsystem ``III'' is dual 
to the $6$--dimensional ones ``I'' and ``II''. Let us note that the signature
$-++++$ of the $5D$ space--time of the subsystem ``III'' corresponds to the
one $--++++$ of the $6D$ signature for the subsystems ``I'' and ``II'' as it
follows from Eqs. (\ref{22}) and (\ref{27}).

Now our goal is to establish the type of the equivalence (\ref{27}) and to 
express the $5D$ field components of the subsystem ``III'' in terms of the 
``physical'' off--shell fields $F$, $f$, $\vec\omega$ and the on--shell 
defined ones which include $H$. At the first time let us calculate the 
scalar field components $e^{\Phi_{III}}$, $e^{2\phi_{III}}$, $G_{III}$ 
and $B_{III}$. Using Eqs. (\ref{1'}), (\ref{22}) and (\ref{27}), one obtains
that
\be \label{28}
e^{\Phi_{III}}=fe^{2\phi_{III}}=-F_{11}, \qquad G_{III}=f(F^{-1})_{22}, 
\ee
where the matrices $F$ and $F^{-1}$ are considered as the block parametrised  
ones with the components ``11'', ``12'', ``21'' and ``22'' of the dimensions 
$1\times 1$, $1\times 2$, $2\times 1$ and $2\times 2$ correspondingly. Then, 
from Eqs. (\ref{22'}) and (\ref{25'}) it follows that
\be \label{31}
B_{III}=-\sigma_2H_1,
\ee
where $\sigma_2$ is the $2\times 2$ antisymmetric matrix with 
$(\sigma_2)_{12}=-1$.

To calculate the vector field components let us consider the motion equation 
corresponding to Eq. (\ref{17}):
\be \label{32}
\nabla\vec J=0.
\ee
From this it follows that it is possible to introduce on--shell the vector
matrix $\vec\Omega$ accordingly
\be \label{33}
\nabla\times\vec\Omega=\vec J.
\ee
Let us parametrise the matrix $\vec\Omega$ in the same manner as $M$; then 
from Eq. (\ref{33}) one obtains that 
$\nabla\times\vec\Omega_{21}=f^{-1}F^{-1}\nabla H$, so
\be \label{33'}
\vec\Omega_{21}=\vec\omega.
\ee
Using the orthogonal coset representation (\ref{20})--(\ref{21}) one can 
also introduce on--shell the vector matrix $\Omega_{BS}$ as
\be \label{33''}
\nabla\times\vec\Omega_{BS}={\cal G}^{-1}\nabla{\cal B}{\cal G}^{-1}.
\ee
Then, using Eqs. (\ref{25'}) and (\ref{27}) after some algebra one concludes 
that
\be \label{33'''}
\vec\omega_m=\frac{1}{2}\epsilon_{mnk}(\vec\Omega_{BS})_{nk}.
\ee
Finally, using Eqs. (\ref{22}), (\ref{22'}), (\ref{33''}) and (\ref{33'''})
for the vector field $\vec A_{1\,\,BS}$ one obtains the following result:
\be \label{33''''}
\vec A_{1\,\,BS}=-\sigma_2
\left(
\ba{cc}
\vec\omega_2\cr 
\vec\omega_3
\ea
\right).
\ee
Now let us denote
\be \label{33'''''}
\vec\Omega_{11}\equiv -\vec\omega^{\star}; 
\ee
then the matrix $\vec\omega^{\star}$ satisfies the differential equation
\be \label{34}
\nabla\times\vec\omega^{\star}=F^{-1}(\nabla F+f^{-1}\nabla H\,H^T).
\ee
From this equation and Eq. (\ref{4}) it follows that 
\be \label{40}
\vec A_{2\,\,BS}=\vec\omega^{\star}_{21},
\ee
where the matrix $\vec\omega^{\star}$ is parametrised in the same manner as 
$F$. Using Eqs. (\ref{2})--(\ref{3}), (\ref{33''''}) and (\ref{40}) one 
immediately obtains the algebraical relations for the mixed and 
the pure $3$--dimensional 
components of the metric and Kalb--Ramond fields; we will not write down them 
here.

It is easy to see that Eqs. (\ref{28}), (\ref{31}), (\ref{33''''}),
(\ref{33'''''})--(\ref{40}) completely define the duality between the
subsystems ``I'' (or ``II'') and ``III''. This duality has the on--shell
nature because the fields $B_{III}$ and $\vec A_{2\,\,BS}$ are defined
in terms of the ``Kaluza--Klein'' ones using the differential relations.
Below we explore the established formulae to construct the duality related
solutions for the subsystems ``I''--``III''.

%%%%%%%%%%%%%%%%%%%%%%%%%%%%%%%%%%%%%%%%%%%%%%%%%%%%%%%%%%%%%%%%%%%%%%%%%

\section{Special solution and its interpretations}

Let us consider the matrix $M$ of the following special form:
\be \label{41}
M=M_0+\lambda \,CC^T,
\ee
where 
\be \label{42}
M_0={\rm diag}\,(-1,-1,1,1),
\ee
$\lambda=\lambda(x^{\mu})$ is the coordinate function and $C$ is the
constant column. Then, the ``matter'' part (\ref{32}) of the motion
equations as well as their Einstein part 
\be \label{43}
R_{3\,\,mn}=\frac{1}{4}{\rm Tr}\left [\left ( \vec J\right)_{\mu}
\left ( \vec J\right)_{\nu}\right ].
\ee
become satisfied if one supposes that
\be \label{44}
\nabla^2\lambda=0,  
\ee
\be \label{44'}
C^TM_0C=0, 
\ee
and $h_{mn}$ corresponds to the flat $3$--space. Also from Eq. (\ref{44'})
it follows that $M=M_{0}e^{\lambda M_0CC^T}$, so ${\rm det}\,M=1$. Let us 
note, that the matrix $M_0$ describes the trivial field configuration for 
the $5D$ bosonic string theory which represents the empty Minkowskian $5D$ 
space--time with the single time--like coordinate. Thus, one obtains the 
correct matrix $M$ signature at least in some vicinity of the spatial 
infinity for any asymptotically trivial solution $\lambda$ of the Laplace 
equation (\ref{44}). 

Let us now introduce the functional vector $\vec\nu$ on--shell of the
solutions of Eq. (\ref{44}) as
\be \label{46}
\nabla\times\vec\nu=\nabla\lambda;
\ee
and the $3\times 3$ matrix $\Sigma={\rm diag}\,(-1\,-1\,1)$. Let us also
parametrise the column $C$ as
\be \label{47}
C=
\left(
\ba{cc}
c_1\cr 
c_2\ea
\right),
\ee
where $c_1$ and $c_2$ are the $3\times 1$ and $1\times 1$ block 
components correspondingly. Then for the matrices $M$, \, 
$M^{-1}=M_0-\lambda M_0CC^TM_0$ and $\vec\Omega=\vec\nu CC^TM_0$ one 
obtains:
\be \label{48}
M=
\left(
\ba{cc}
\Sigma+\lambda c_1c_1^T&\lambda c_2c_1\cr 
\lambda c_2c_1^T&1+\lambda c_2^2
\ea
\right)
\quad 
M^{-1}=
\left(
\ba{cc}
\Sigma-\lambda \Sigma c_1c_1^T\Sigma &-\lambda c_2\Sigma c_1\cr 
-\lambda c_2c_1^T\Sigma &1-\lambda c_2^2
\ea
\right),
\ee
\be \label{49}
\vec\Omega=\vec\nu
\left(
\ba{cc}
c_1c_1^T\Sigma &c_2c_1\cr 
c_2c_1^T\Sigma &c_2^2
\ea
\right).
\ee
Then, using Eqs. (\ref{18}) and (\ref{48}) one calculates the fields
\be \label{50}
f^{-1}&=&1-\lambda c_2^2, \quad H=\frac{\lambda c_2\Sigma c_1}
{1-\lambda c_2^2},\nonumber\\ 
F&=&\Sigma-\frac{\lambda\Sigma c_1c_1^T\Sigma}{1-\lambda c_2^2},
\quad
F^{-1}=\Sigma+\lambda c_1c_1^T,
\ee
and also $\vec\omega=c_2c_1\vec\nu$.

Now we are ready to write down the corresponding solutions for the 
subsystems ``I''--``III''. Actually, from Eqs. (\ref{14})--(\ref{15}) 
it immediately follows that  the pure Kaluza--Klein solution ``I''
reads:
\be \label{52}
ds^2_{I}=(dy+c_2c_1\vec\nu d\vec x)^T\left (
\Sigma-\frac{\lambda\Sigma c_1c_1^T\Sigma}{1-\lambda c_2^2}
\right )(dy+c_2c_1\vec\nu d\vec x)+(1-\lambda c_2^2)
h_{\mu\nu}dx^{\mu}dx^{\nu}.
\ee
Then, the solution for the subsystem ``II'' has the form
\be \label{53}
ds^2_{II}&=&dy^T\left (
\Sigma+\lambda c_1c_1^T\right )dy+(1-\lambda c_2^2)
h_{\mu\nu}dx^{\mu}dx^{\nu},
\nonumber\\
e^{\Phi_{II}}&=&1-\lambda c_2^2,\quad B_{II\,\,m,d+\mu}=
c_2(c_1)_m\nu_{\mu}.
\ee
Finally, for the subsystem ``III'' the developed ``technology'' 
(see Eqs. (\ref{28}), (\ref{31}), (\ref{33''''}), 
(\ref{33'''''})--(\ref{40}) gives the following result:
\be \label{55}
ds^2_{III}=-(dy-c_2\sigma_2c_1^{''}\vec\nu d\vec x)^T\left (
\frac{\sigma_3-\lambda c_1''c_1^{''\,\,^T}}{1-\lambda c_2^2}
\right )(dy-c_2\sigma_2c_1^{''}\vec\nu d\vec x)+(1-\lambda c_1^{'\,2})
h_{\mu\nu}dx^{\mu}dx^{\nu},
\nonumber
\ee
\be 
e^{\Phi_{III}}=\frac{1-\lambda c_1^{'\,2}}{1-\lambda c_2^2},  \quad 
B=\frac{\lambda c_2c_1^{'}}{1-\lambda c_2^2}\sigma_2, \quad
B_{m,d+\mu}=\frac{c_1^{'}(c_1^{''})_m\nu_{\mu}}{1-\lambda c_2^2}.
\ee
Here $\sigma_3={\rm diag}\,(1,-1)$, whereas $c_1^{'}$ and $c_1^{''}$ are 
the $1\times 1$ and $2\times 1$ block components of the $3\times 1$ column 
$c_1$:
\be
c_1=
\left(
\ba{cc}
c_1^{'}\cr 
c_1^{''}\ea
\right).
\ee

It is easy to check that $B_{d+\mu,\,d+\nu}=0$ for the both subsystems 
``II'' and ``III'' because in these cases the matrix vector functions 
$\vec A_1$ and $\vec A_2$ are the products of the corresponding constant 
matrices and the single functional vector field $\vec\nu$. The solutions
(\ref{52})--(\ref{55}) describe the extremal field configurations related 
to the single real harmonic function. In the Einstein--Maxwell theory 
solutions of this type form the Majumdar--Papapetrou solution class 
\cite{extr};
in the superstring theory they contains the BPS--saturated solutions 
\cite{y}, \cite{bps1}, \cite{bps2}. Taking the function $\lambda$ as the 
sum of the Coulumb--like 
terms, one obtains for all these theories as well as for our subsystems 
``I''--``III'' the field configurations which describe a set of the arbitrary 
located point--like interacting sources. 

%%%%%%%%%%%%%%%%%%%%%%%%%%%%%%%%%%%%%%%%%%%%%%%%%%%%%%%%%%%%%%%%%%%%%%%%

\section*{Acknowledgments}
This work was supported by RFBR grant ${\rm N^{0}} 
\,\, 00\,02\,17135$. O.V.K.
thanks Instituto de Fisica y Matematica for facilities and hospitality
during his stay at Morelia, Mexico when this work was partially completed. 
%%%%%%%%%%%%%%%%%%%%%%%%%%%%%%%%%%%%%%%%%%%%%%%%%%%%%%%%%%%%%%%%%%%%%%%%%%%%%

\section{Conclusion}

From the formal point of view, the dual theories are equivalent and any result
obtained for the one theory can be immediately rewritten as the result for the 
another theory which belongs to the same ``dual multiplet''. In the study of 
the dual theories it is useful to find the most convenient representation and
to use it for the problem solution; the last step is related to the translation
of the obtained results into the language of the original physical variables. 

In this paper we have studied the ``dual triplet'' which consists of the 
$6$--dimensional complete Kaluza--Klein theory, $6$--dimensional truncated
bosonic string theory and $5$--dimensional complete bosonic string theory
compactified to three dimensions on a torus. In this case the most 
convenient representation is given by the $6D$ Kaluza--Klein theory chiral 
formalism. We have established the correspondence formulae (the duality) and 
have constructed the class of the extremal solutions which are related to 
the single real harmonic function. In fact it is possible to generalize this 
class to the case of two real harmonic functions; the corresponding 
solutions will have the Israel--Wilson--Perjes type \cite{iwp}. 
Using the presented general formulae one can map the wide classes 
of known solutions of the $6$--dimensional Kaluza--Klein theory into the
$5$-- and $6$--dimensional solutions of the bosonic string theory. However, 
as it follows from our consideration, one must start from the 
$6$--dimensional Kaluza--Klein theory with two time--like coordinates
to obtain the correct signature in $5$ dimensions. Here we will not discuss 
this formally evident but physically surprising fact.

Our dual theories (the subsytems ``I''--``III'') form the special cases of 
the low--energy bosonic string theory. This means that the bosonic 
string theory (in $7$ dimensions at least in view of two necessary time--like 
dimensions for the $6D$ subsystems) can be considered as some known 
``generating'' theory for our ``dual multiplet''. The similar situation 
takes place in the superstring theory where it is known the set of five 
consistent concrete theories \cite{gsw} which are related by the web of 
string dualities. In this case the ``generating'' theory is named as 
``M--theory''; now this theory is under construction \cite{kir}.
%%%%%%%%%%%%%%%%%%%%%%%%%%%%%%%%%%%%%%%%%%%%%%%%%%%%%%%%%%%%%%%%%%%%%%%%%%%%%%%

\end{document}